# Vertical and temporal $H_3^+$ structure at the auroral footprint of Io

A. Mura[1], A. Moirano[1,2], V. Hue[3], C. Castagnoli[1,4,5], A. Migliorini[1], A. Altieri[1], A. Adriani[1], A. Cicchetti[1], C. Plainaki[6], G. Piccioni[1], R. Noschese[1], G. Sindoni[6], R. Sordini[1].

1. Institute for Space Astrophysics and Planetology, National Institute for Astrophysics, Roma, Italy

2 Laboratory for Planetary and Atmospheric Physics, Space Sciences, Technologies and Astrophysical Research Institute, University of Liège, Liège, Belgium

3 Aix-Marseille Université, CNRS, CNES, Institut Origines, LAM, Marseille, France

4 University of Rome "Tor Vergata", Roma, Italy

5 ISAC-CNR, Bologna, Italy

6 Agenzia Spaziale Italiana, Rome, Italy

**Abstract.** We report the first observation of the vertical and temporal structure of the $H_3^+$ emission at the auroral footprint of Io, as observed by Juno/JIRAM. The brightness vertical profile shows a maximum at 600 km above 1 bar, with no apparent difference between the Main Alfvén Wing spot emission and the tail of the footprint. This observation is more compatible with a broadband energy distribution of the precipitating electrons, than a monoenergetic one. The temporal profile of $H_3^+$ column density has been observed after the passage of the MAW and shows a hyperbolic decrease. A model of $H_3^+$ decay is proposed, which takes into account the second-order kinetic of dissociative recombination of $H_3^+$ ions with electrons. The model is found to be in very good agreement with Juno observation. The conversion factor from radiance to column density has been derived, as well as the half-life for $H_3^+$, which is not constant but inversely proportional to the $H_3^+$ column density. This explains the wide range of $H_3^+$ lifetimes proposed before.

Key points:

Vertical structure of the auroral footprint of Io

Broadband electron energy distribution is compatible with the observations.

$H_3^+$ density decays in a hyperbolic way, lifetime is not defined.

Implications for UV-IR comparisons.

# 1. Introduction

The infrared emission spectrum of $H_3^+$ was first discovered by Oka (1980); a few years later, Drossart et al. (1989) identified such emission from the Jupiter atmosphere. They proposed that the formation of $H_3^+$ was due to the ionization, from electron precipitation and photoionization, of neutral $H_2$ molecules which quickly react with neutral $H_2$ to form $H_3^+$ and H. The reaction is highly exothermic so that a considerable amount of heat is deposited in the atmosphere with this reaction. Drossart et al. (1989) also proposed that this process could occur in the upper atmosphere of Jupiter ($h > 500$ km) because, at that altitude, $H_3^+$ emission is very intense, while the radiation from the deeper Jupiter's atmosphere is shielded by the more abundant methane layer, at ~200 km (Connerney et al., 2000). Because of the correlation with electron precipitation, Connerney et al., (2000) proposed to use such emission as a diagnostic tool to study the magnetosphere of Jupiter, and, for this purpose, the NASA mission Juno to Jupiter (Bolton et al., 2017) is equipped with a dedicated instrument, the Jovian InfraRed Auroral Mapper (JIRAM) (Adriani et al., 2008, 2017a). The JIRAM scientific goals are to explore the Jovian aurorae and the planet's atmospheric structure, dynamics and composition. JIRAM images and spectra of the $H_3^+$ emission have showed the auroral regions of Jupiter with unprecedented detail, observing the northern and southern auroras two hours apart (Adriani et al., 2017b; Dinelli et al., 2017; Moriconi et al., 2017; Mura et al., 2017). Outside of the main oval, auroral emissions coming from the electromagnetic interactions between Jupiter and its moons (particularly Io) are also detected by JIRAM (Mura et al., 2018; Moirano et al., 2021, 2023). Compelling features of Jupiter's aurora, and absent at Earth, these emissions are located at the magnetic footpoints of the Galilean moons Io, Europa, and Ganymede (Connerney et al. 1993, Clarke et al. 1996, Zarka 2004). They stem from the interactions between the moons and the corotating magnetized plasma that fills the extended space environment, or magnetosphere, of this giant planet. Several processes are proposed to explain these auroral features. One theory evokes plasma (Alfvén) waves, that travel from the moon to Jupiter northern and southern high-latitude region, along the magnetic field, (Acuna et al. 1981, Belcher et al. 1981, Neubauer 1980, Kivelson et al. 2004), bouncing between hemispheres (Gurnett and Goertz, 1981), and accelerating electrons as they proceed both toward and away from Jupiter along the magnetic field lines (e.g., Hess et al. 2010). Electrons precipitating into the Jovian atmosphere produce hydrogen emissions associated with the moon-magnetosphere interaction, equatorward of the main auroral arc (Bonfond et al. 2017). Juno infrared and UV observations of the moon-generated auroral features (Mura et al., 2018, Szalay et al. 2018; Moirano et al., 2021, 2023; Hue et al., 2022, 2023) show much more structure than anticipated from Earth-based observations. By using JIRAM images, Mura et al., (2018) found that the footprint feature has a quite complex morphology with multiple small dots arranged in a zig-zag pattern, like a von Kármán vortex street (von Kármán, 1911), which eventually turns into a turbulent tail. They also noted that the timescale for the decay of the intense $H_3^+$ emission of these features was shorter (~60 s) than the typical timescale for $H_3^+$ destruction (~1000 s, Stallard et al., 2002) proposed earlier.

Knowing the decay rate of $H_3^+$ is crucial for understanding the comparative analysis of UV and IR images of Jupiter's aurora, as it helps to accurately interpret the temporal and spatial differences in the two bands (Gerard et al., 2020). As noted by Tao et al. in 2011, the reaction rate depends on the square of the density. In such reactions, known as second-order chemical kinetics, the classic exponential decay does not apply; therefore, as it will be shown in this study, the definition of an $H_3^+$ lifetime can be misleading, as it depends on $H_3^+$ density. Density varies with altitude and reaches a maximum around 600 or 700 km above 1 bar (Migliorini et al., 2023). Consequently, the $H_3^+$ decay will follow a complex pattern, and an accurate modeling of the vertical profile is necessary to obtain precise results. In this study we will first derive the vertical profile of $H_3^+$ emission, and then apply it to a model of $H_3^+$ decay. Finally, we will compare the model with data.

In Section 2, we describe the vertical structure of Io's auroral footprint. Section 3 presents data on the temporal variability of secondary spots. In Section 4, we outline the model and conduct the analysis. A summary and conclusions are provided in Section 5. The instrument and dataset are detailed in Appendix A1 and A2.

## 2. Vertical profile

In Figure 1 we present the first detailed image of the vertical profile of Io's auroral footprint emission, observed in the north. Previously, footprint images taken at the limb were produced, including those acquired by the Hubble Space Telescope (HST). However, none were captured at this level of detail, which allows for an accurate reconstruction of the vertical profile. Figure 1 was created using 32 images taken over the northern Io auroral footprint during orbit 31, on Dec. 30, 2012, from 21:05 to 21:25 (see Appendix A2 for a table of the observations). These images represent L-band radiance, which is integrated from 3.3 to 3.6 μm; the images were projected onto a plane defined by three points: the origin at the center of Jupiter; the estimated location of Io's footprint at the start of the observation; the estimated location of Io's footprint at the end of the observation. By assuming that the $H_3^+$ emission originates from this plane, or close to it, we determine the height relative to the 1-bar pressure level. Since we assume the footprint emission is localized, this approach avoids the need to develop an "onion-peeling" model. A minor source of error arises from possible deviation of the actual location of the footprint with respect to the estimated one. Here we use the model by Connerney et al. (2018) to perform the estimation of the footprint position.

The x- and y-axes of the figure represent distances in kilometers relative to an arbitrary point. The four curved lines in the figure represent Jupiter's 1-bar pressure level and three other levels at 250 km, 500 km, and 750 km, respectively. Both the MAW (Main Alfvén Wing) and the RAW (Reflected Alfvén Wing, Bonfond et al., 2008, 2009, 2010) spots are clearly visible in the mosaic and are indicated by the red arrows. At the same location of the RAW, a TEB (Trans-hemispheric Electron Beam, Bonfond et al., 2008) can also contribute to the emission (the TEB is sometimes ahead of the MAW, sometimes following it; in this case, the predicted location is in fact the same as that of the RAW).

Panel B zooms in on the intensity of radiance as a function of altitude, between 0 and 750 km, with the highest intensity roughly corresponding to the main spot. The radiance intensity, originating from $H_3^+$, peaks at approximately 600 km; the height of the column of emitting $H_3^+$ is approximately 100 or 200 km.

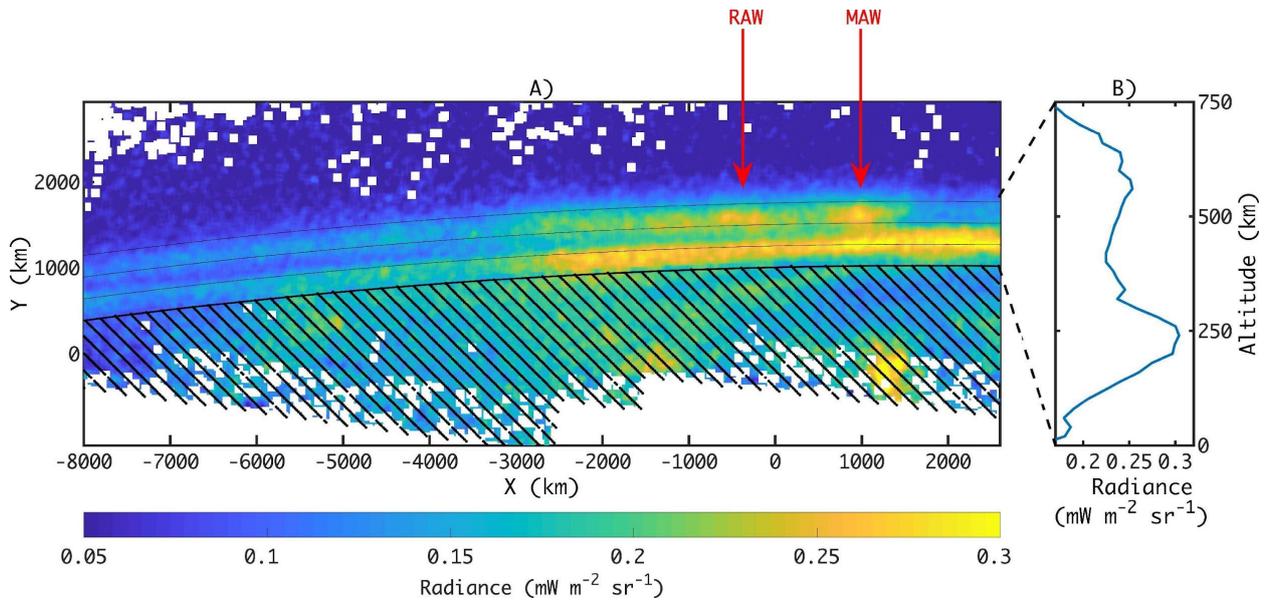

**Figure 1.** Panel A: L-band (3.3–3.6 µm) emission from Io's auroral footprint (north), created using 32 images from orbit 31 (Dec. 30, 2012). The x- and y-axes show distances in kilometers (arbitrary origin of axis), with four curved lines indicating Jupiter's 1-bar pressure level and altitudes of 250, 500, and 700 km. Radiance intensity, mainly from $H_3^+$, peaks at approximately 600 km, with an emitting column 100-200 km high. Methane emission is visible near 250 km. MAW (Main Alfvén Wing) and RAW (Reflected Alfvén Wing) spots are indicated by the red arrows. Close to the location of the RAW, a TEB can also contribute to the emission. The predicted location of the MAW is ahead of the observed MAW by about 1000 km. Panel B: zoom on radiance intensity between 0 and 750 km, calculated on a column corresponding to the main spot.

At 3.3 µm, mixed with the substantial $H_3^+$ emission, weaker emission due to methane has been observed in the auroral emission (Adriani et al. 2017; Dinelli et al., 2017; Moriconi et al. 2017) when observed vertically. The emission we observe near 250 km can be attributed to methane, as shown by Migliorini et al. (2023). The emission from $CH_4$ looks more intense than the one coming from $H_3^+$, just because it is seen tangentially and not from above as usual, i.e. the line of sight is much longer for $CH_4$. In any case, this emission is separable from the $H_3^+$ emission. There may also be a background auroral emission that cannot be eliminated or removed from this figure. In summary, it is quite clear that the dominant emission in the region between 500 and 750 km is from $H_3^+$.

The vertical profile of $H_3^+$ radiance can, in principle, provide insights into the energy spectrum of the precipitating electrons. According to Tao et al. (2011), the production rate of $H_3^+$ from a monoenergetic beam of electrons at 10 keV peaks at approximately 600 km, which coincides with the observed radiance peak. Achieving a more accurate fit (i.e. using a complete electron spectrum) is challenging, particularly given that the production rate does not directly scale to the density profile. Nonetheless, it is worth noting that, in this specific case, the vertical profile of $H_3^+$ radiance is inconsistent with electrons of either 1 keV or 100 keV (see Figure 3 in Tao et al., 2011).

## 3. Time evolution of $H_3^+$ emission from the footprint of Io.

Mura et al. (2018) showed that the auroral footprint of Io (IFP) consists not only of the well-known MAW, RAW, and TEB but also of a series of secondary spots that follow the MAW in Io's footprint ("sub-dots"). Moirano et al. (2021) analyzed these sub-dots and demonstrated that they are stationary in the SIII reference frame (i.e. in the Jovian ionosphere). The cause of these spots remains unclear; as noted by Schlegel et al. (2022), multiple physical interpretations are possible.

Nonetheless, the most straightforward interpretation is that these spots are formed following the passage of the MAW, due to transient electron precipitation (the physical reason why this electron precipitation may be pulsating remains unclear, but is not relevant for this study). After the passage of the MAW we observe a stationary but dimming sub-dot, due to the disappearance of $H_3^+$ caused by recombination with electrons.

With this premise, and acknowledging that we cannot entirely rule out subsequent electron precipitation, we produce maps of the L-band radiance of the south IFP at 30 s time steps; we also define a region of interest (ROI) of 200 × 200 km over one of the sub-dots. We then calculate the average radiance within the ROI, ensuring that a single stable sub-dot remains within this region.

Moirano et al. (2021) presented a very long sequence of 16 images (i.e. ~ 8 minutes), as shown in Figure 3, in which sub-dots are stationary, and they excluded the possibility of sub-dot motion that occurs on timescales in between the observations (every 30 s). However, there are at least two reasons why we prefer to avoid analyzing such long sequences. The first reason is that the RAW (or sometimes a TEB), following the MAW, would eventually enter the region of interest, compromising the analysis, which relies on the absence of any other electron precipitation in the area during the study. The second reason is due to ionospheric winds, which can reach speeds of the order of 1 km/s at the height of $H_3^+$ (Maillard et al., 1999; Rego et al., 1999; Stallard et al., 2003; Johnson et al., 2017). To avoid potential blurring caused by these winds, it is preferable to limit the analysis to sequences shorter than 200 seconds (that is, the size of the ROI divided by the speed of winds).

The JIRAM database contains several useful examples; we focus on the two best: one from orbit 13 and the other from orbit 26. The first example is illustrated in Figure 2, with further details provided in the appendix. Seven orthographic maps of L-band radiance, detected by JIRAM and projected to a level of

650 km, are shown. The radiance is corrected for the emission angle (*e*) with a cosine law - that is, we divide by cos(*e*) to take into account the longer line of sight for slant observations. The ROI is represented by the red square; the mean value of the radiance in the ROI is given in table A2. A second dataset, obtained from orbit 26, provides eight useful images and it is described in table A3.

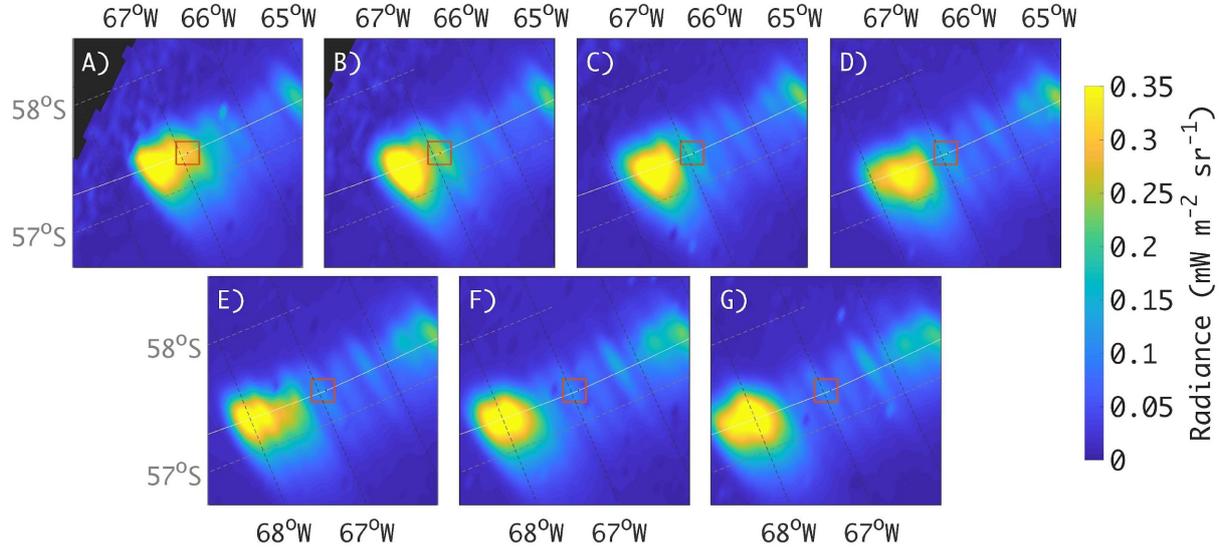

**Figure 2**. Seven observations (radiance maps in the L band) of the south Io Footprint (IFP) taken during Juno's orbit 13 (from 07:12:30 to 07:15:30 on 2018-05-24). The red square is the Region Of Interest (ROI) used to calculate the mean radiance from a fixed ionospheric column. The white line is the trajectory of the IFP predicted by the model of Connerney et al. (2018). The dashed lines are meridians (67° and 68° W) and parallels (57° and 58° S).

## 4. Theory and modelling

### 4.1. Formation and destruction of $H_3^+$

$H_3^+$ is predominantly formed at altitudes ranging from 500 to 1000 km, with its maximum concentration occurring around 600 to 700 km above the 1-bar pressure level. In the model proposed by Tao et al. (2011), this peak is calculated as a function of the energy of the electron beam (*E*), and it is at 600 km for *E*= ~10 keV (refer to their figure 6b). Other studies have observed the maximum $H_3^+$ emission at approximately 700-900 km and 680-950 km for $H^3_+$ overtone and hot overtone emissions in the northern auroral region (Lystrup et al., 2008; Uno et al., 2014). Watanabe et al. (2018), who utilized the same model as Tao et al. (2011), found that the peak altitudes for electrons with energies of 1, 10, 30, 100, and 300 keV are 1400, 770, 600, 560, and 560 km above the 1-bar level, respectively. In non-auroral, mid-latitude regions, $H_3^+$ emission occurs at altitudes between 300 and 500 km above the 1-bar level (Stallard et al., 2015, using Cassini/VIMS data; Migliorini et al., 2019, using Juno/JIRAM data). Additionally, Migliorini et al. (2023) present limb observations from JIRAM data, reporting detached infrared emissions from both $H_3^+$ and methane. This finding supports our assumption that $H^3_+$ emission can be effectively isolated in JIRAM data.

Finally, in the previous section, we estimate the $H_3^+$ peak emission altitude in the specific auroral footprint region (which is the most appropriate quantity here) as ~600 km with good accuracy. At these altitudes, in any case, vibrational local thermal equilibrium (LTE) can generally be assumed (Melin et al., 2005, Dinelli et al., 2017). This means that the vibrational temperature closely matches that of the thermosphere/ionosphere (i.e., Lam et al., 1997; Miller et al., 1997). The peak of auroral production of $H_3^+$ is observed at 600-650 Km, where the atmosphere consists primarily of neutral $H_2$ (Achilleos et al., 1998; Tao et al., 2011). By comparing in-between collision time and the typical time before emission from $H_2$ and $H_3^+$, Stallard et al., (2002) concluded that LTE is a reasonable assumption for the region where $H_3^+$ emits. Similarly, Tao et al. (2011) found that LTE becomes negligible below 1000 km.

The interaction of charged particle precipitation with Jupiter's $H_2$-dominated atmosphere leads to auroral emission from $H_2$ and H in the FUV (Clarke et al. 1996, 1998; Badman et al., ), which is another diagnostic tool for the magnetosphere of Jupiter. There is also some emission in the visible range (Ingersoll et al., 1998; Vasavada et al. 1999). Auroral UV emission reflects the immediate energy input from impacting electrons, while IR emission shows the atmospheric response through heating and energy deposition (Gerard et al., 2018). The neutral atmosphere quickly thermalizes the $H_3^+$ ions after their formation. If LTE assumption is correct then it is possible to infer the thermosphere temperature by modelling the relative strength of $H_3^+$ roto-vibrational lines (Lam et al., 1997; Stallard et al., 2002, Dinelli et al., 1992, 1995, 2017). In summary, $H_3^+$ IR emission can be used to both derive the atmospheric temperature and integrated column, thus allowing mapping the ion distribution (Dinelli et al., 2017).

The $H_3^+$ is formed by the reaction:

$$H_2 + H_2^+ \rightarrow H_3^+ + H$$

the reaction rate is $2\times10^{-9}$ cm$^3$ s$^{-1}$ (Perry et al., 1999) and since the $H_2$ density below 1000 km exceeds $10^{10}$ cm$^{-3}$, any $H_2^+$ quickly results in a $H_3^+$ molecule. The new $H_3^+$ is in an excited state and immediately decays with the production of IR radiation. The excited states are continuously refilled via collisions, so that $H_3^+$ emits in the IR until it is destroyed. The recombination with hydrocarbons is the main destruction channel at low altitudes (~less than 500 km), while the electron recombination dominates above the methane homopause:

$$H_3^+ + e \rightarrow H + H + H \text{ or } (H_2 + H)$$

$$H_3^+ + CH_4 \rightarrow CH_5^+ + H_2$$

## 4.2. Decay rate of $H_3^+$

The density and temperature of auroral electrons can be assumed to be the same as those of $H_3^+$ (Tao et al., 2011; Waite et al., 1983). Therefore, after the passage of the electron precipitation associated with the MAW spot over a given area of Jupiter's atmosphere, the local concentration of $H_3^+$ decays according to the following differential equation:

$$\frac{dn_{H_3^+}}{dt} = -k\, n_{H_3^+}^2 \tag{1}$$

(see also Eq. 9 in Tao et al., 2011). The value of $k$ for the destruction rate has been repeatedly questioned (see Miller et al., 2020, and references therein); a critical review of the experiments that measured $k$ and its uncertainty is provided by Larsson et al. (2008) and Larsson (2012). Here we adopt the formula given by Sundstrom et al. (1994):

$$k = 1.15 \times 10^{-7} \left(\frac{300}{T_e}\right)^{0.65} \tag{2}$$

$k$ is given in cm$^3$ s$^{-1}$, and $T_e$ is the electron temperature, in K. We note that this same value used by Tao et al. (2011), and hence by Watanabe et al. (2018), since their cited reference (Perry et al., 1999), inherits the work of Sundstrom et al. (1994); the consistency between published studies justifies the choice of this value.

At the altitude of the IFP emission observed in Fig. 1, we can neglect the destruction of $H_3^+$ by CH$_4$ since the rate (2.4×10$^{-9}$ cm$^3$ s$^{-1}$; Huntress, 1975; Atreya and Donahue, 1975; Huntress, 1977) is ten times lower than $k$, so that this reaction is only effective when $n_{CH_4}$ is larger than $n_{H_3^+}$, which is not the case at $h > 500$ km (see a review of $n_{CH_4}$ vertical profiles in Sanchez-Lopez et al., 2022).

The solution of the differential equation (1) can be easily obtained by separation of variables and is:

$$n_{H_3^+} = \frac{1}{c + kt} \tag{3}$$

for $t=0$, $n_{H_3^+} = 1/c$; hence, $c$ is a parameter that has the dimensions of cm$^3$ and is the inverse of the $H_3^+$ density at $t=0$. The process we are describing, where the rate of change is proportional to the square of the concentration, is known as "second-order kinetics" (or a "second-order reaction" in chemistry) and is common in reactions where two reactant molecules need to collide for the reaction to occur, and the reaction rate depends on the concentration of both. The term "hyperbolic decay" can be used to describe how the concentration decreases over time for second-order processes, as opposed to the exponential decay seen in first-order kinetics. Second-order kinetics are commonly seen in bimolecular reactions, such as two molecules of the same or different species reacting together.

To obtain the radiance in the L band, it is sufficient to integrate $n_{H_3^+}$ along the line of sight, keeping in mind that $n_{H_3^+}$ is not uniform and should be calculated at different locations. A factor $f$ transforms the $H_3^+$ column density into L-band radiance ($R_L$):

$$R_L = f \int n_{H_3^+} \, dl \qquad (4)$$

where $f$ depends on the Einstein coefficients for spontaneous emission associated with the $H_3^+$ transitions in the L-band and the thermospheric temperature. We intentionally leave it free in the best-fit that we will perform in the next section, and we will then check if we obtain realistic values for $f$ and for the column density.

### 4.3. Model and data analysis.

The model to reproduce the decay of the radiance observed in Figure 2 is obtained by assuming that, at the initial instant $t=0$, the vertical profile of $H_3^+$ follows a Gaussian distribution. This distribution peaks at 600 km with a sigma of 100 km, consistent with the data presented in Figure 1, panel B. The Gaussian form is described by:

$$n_{H_3^+}(t=0) = n_p \, e^{-\frac{(h-600)^2}{2 \times 100^2}} \qquad (5)$$

where $h$ represents the altitude above the 1-bar pressure level, measured in km, and $n_p$ denotes the peak density at 600 km and at $t=0$. By combining this expression (Eq. 5) with the previous ones, we obtain the $H_3^+$ density as a function of time. Further, using this result in Eq. 4 allows us to estimate the radiance generated by the column, directed upward.

For simplicity, $k$ is assumed uniform between 500 and 800 km (using the temperatures from Tao et al., 2011, for these altitudes, which are approximately 700 K at 500 km and 1000 K at 800 km, gives very similar values for $k$ in Eq. 2: 6.6 and 5.2×10$^{-8}$ cm³ s$^{-1}$). We adopt an intermediate value of 6.5×10$^{-8}$ cm³ s$^{-1}$ at 900 K, a temperature for $H_3^+$ emission as in Dinelli et al. (2017).

The free parameters in the model are thus reduced to two: $n_p$ and the factor $f$. We fit these parameters to the data extracted from Figure 2 to evaluate the model's performance. To ensure robustness, we also perform a fit using a Markov Chain Monte Carlo (MCMC) method. In Figure 3, we present the results for orbits 13 (panel A) and 26 (panel B). The best-fit results are shown as dashed lines, while the MCMC density plots are represented in gray. Table 1 provides the fitted values.

**Table 1**

| Orbit | $n_p$ (cm$^{-3}$) | $f$ ($10^{-18}$ W m$^{-2}$s$^{-1}$ / cm$^{-2}$) | $\chi^2$ mW m$^{-2}$s$^{-1}$ |
|---|---|---|---|
| 13 | $3.5 \times 10^5$ | 40 | $7 \times 10^{-3}$ |
| 26 | $1.2 \times 10^5$ | 80 | $1.7 \times 10^{-3}$ |

We observe good agreement between the data and the model. The $\chi^2$ values for orbits 13 and 26 are $7 \times 10^{-3}$ and $1.7 \times 10^{-3}$ mW m$^{-2}$s$^{-1}$, respectively. Notably, an observer might not easily distinguish between this decay profile and an exponential one at first glance. However, a best-fit with an exponential function yields significantly worse $\chi^2$ values: $15 \times 10^{-3}$ and $2.7 \times 10^{-3}$ mW m$^{-2}$s$^{-1}$, respectively.

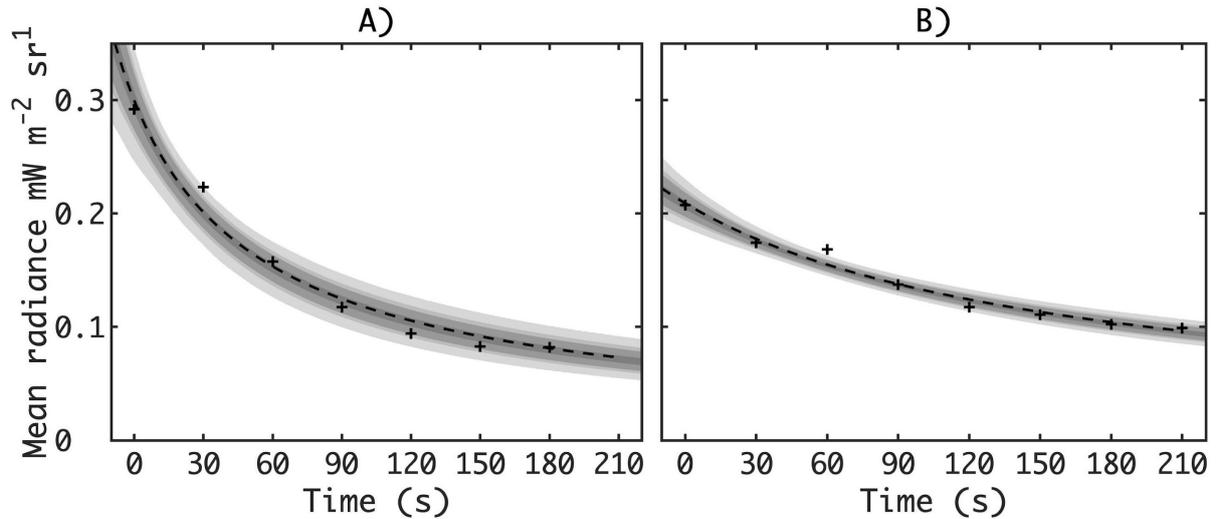

**Figure 3**. Temporal variability for the mean radiance within the ROI (panel A: orbit 13; panel B: orbit 26). The dashed line represents the best-fit function. The gray regions represent the density plot from the MCMC model, which is representative of the model uncertainty.

Figure 4 presents the corner plot for orbit 13, illustrating that the parameters are well constrained. Although there is an evident anticorrelation between $n_p$ and $f$, this does not significantly impact the results, as the uncertainties are reasonably small. Nevertheless, the model is not meant to serve as an absolute method for calculating the column density. The approximately 50% uncertainty in $f$ is acceptable.

This anticorrelation resembles what was observed by Dinelli et al. (2017) or Adriani et al. (2017) between temperature and column density. What is most remarkable, however, is that the value of $f$ derived solely from the temporal behavior of the radiance—without any additional information beyond electron recombination—aligns very well with values reported in the literature. For instance, Dinelli et al. (2017) report that a radiance of $1\times10^{-4}$ W m$^{-2}$ sr$^{-1}$ corresponds to a column density of $2\times10^{12}$ cm$^{-2}$, while a radiance of $1.5\times10^{-4}$ W m$^{-2}$ sr$^{-1}$ corresponds to $3\times10^{12}$ cm$^{-2}$. This translates to a typical ratio ($f$) of approximately $50\times10^{-18}$ W m$^{-2}$ s$^{-1}$ / cm$^{-2}$, which matches our results (ranging from 40 to 80 $\times10^{-18}$).

The variation in $f$ between the two datasets may be due to statistical fluctuations, given the inherent noise in the data. Moreover, orbit 13 has an emission angle of 52 degrees, while orbit 26 has 33 degrees, potentially introducing an effect related to sub-optimal emission angle correction.

In conclusion, the model shows good agreement with the data, demonstrating that it effectively captures the decay behavior of the radiance.

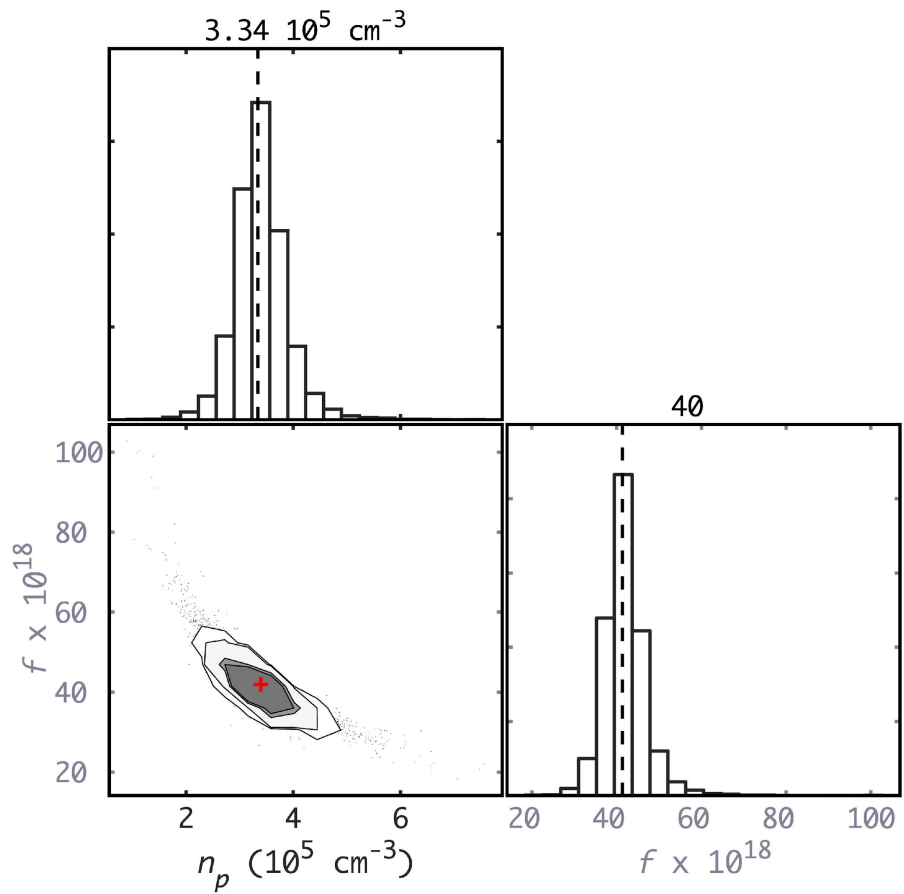

**Figure 4.** Corner plot for the MCMC model (orbit 13 only). The free parameters are the peak density at t=0 ($n_p$) and the factor $f$ (units of W m$^{-2}$s$^{-1}$ / cm$^{-2}$) that converts from column density to radiance. The numbers above the plots are the median values from the distribution.

By taking the model functions in Figure 3 (dashed lines) it is possible to calculate the half-life as a function of the column density. The result is given in the Appendix (Figure A1). The relationship is an almost perfect hyperbole, and it can be fitted with this formula:

$$T_{\frac{1}{2}} = \frac{5 \times 10^{14}}{CD} \tag{6}$$

where $T_{1/2}$ is the half-life in seconds, and CD the column density in cm$^{-2}$.

## 5. Discussion and conclusions

The first notable result of this study is the derivation of the vertical profile of the Io footprints auroral emission in the infrared. The peak altitude of $H_3^+$ emissions is lower than that of UV emissions, and also lower than generally assumed. By comparing the UV and IR emission altitude predicted by a monoenergetic electron distribution precipitating into a model atmosphere (Tao et al., 2011), we see that for energy greater than 1 keV, the UV emission peaks below the IR emission. Indeed, the hydrocarbons at low altitudes destroy the $H_3^+$ before it can radiate, but in principle, it is still possible to see UV light near 160 nm due to the de-excitation of the hydrogen (there is negligible UV absorption by hydrocarbons around that wavelength; Gustin et al. (2016). Conversely, for energies below 1 keV, the UV emission occurs above the IR emission. According to this IR versus UV comparison, we may suggest that the bulk of the precipitating electrons in the MAW has an energy below 1 keV. This would be in qualitative agreement with Juno's particle measurements, which reported the peak of the energy distribution between 0.1 and 1 keV (Szalay et al., 2020). We note, however, that one 1 keV electron beam, in the same model reported above (Tao et al., 2011), peaks at much higher altitudes (1000 km), and 1 keV (or lower) electrons do not contribute significantly to the $H_3^+$ emission at 600-660 km; 10 or even 100 keV electrons are more in agreement with the 600 km value derived from Fig. 1. We therefore conclude that, since the $H_3^+$ vertical profile is largely governed by the destruction rate, assessing the energy spectrum of the precipitating electrons is a quite difficult task. We propose that the solution to this discrepancy could be that mono-energetic beams are not a suitable assumption to model the electron precipitation of the IFP. Indeed, Juno's particle measurements reported broadband distributions (Szalay et al., 2020), which can lead to different vertical brightness profiles from the monoenergetic case (Benmahi et al., 2024). Previous observations of the IFP were taken from Earth using the Hubble Space Telescope (Bonfond et al., 2009, Bonfond, 2010). A sequence of observations of the IFP in 2007 revealed that the UV brightness vertical profile of the MAW spot can be approximated by a Chapman function that peaks at about 800 km and with a full width at half maximum of ~850 km. The UV emission of the footprint tail peaks at about 900 km, although large variations between 700 and 1100 km have been reported. A similar altitude has been reported to support particle measurement connected to Io's magnetic shell (Szalay et al., 2018) using observations of the IFP tail by Juno-UVS. The vertical extent of the IFP UV brightness suggests that the electron energy distribution is predominantly broadband. The best match seems to be achieved by a *K*-distribution with a relatively low characteristic energy of 70 eV, mean energy of ~1 keV and *K*=2.3 (Bonfond et al., 2009). As suggested by theoretical efforts (Damiano et al., 2019; Hess et al., 2010), such energy distribution is associated with precipitating electrons that are accelerated by wave-particle interaction with Alfvén waves. Juno's magnetometer (Connerney et al., 2017) measured intense Alfvénic activity during the crossing of Io's magnetic shell (Gershman et al., 2019; Sulaiman et al., 2020), and Juno's particle detector JADE-E (McComas et al., 2017) recorded broadband energy spectra between 0.05 keV and 30 keV at various distance from the MAW spot along the footprint tail (Szalay et al., 2018; Szalay et al., 2020). Furthermore, Benmahi et al. (2024) showed that, in the main jovian aurora, the assumption of monoenergetic electron energy distribution leads to quantitatively different results from a broadband distribution, when the UV auroral emission is used to derive the corresponding precipitating electron energy. The precipitation in the main emission is in general far more energetic than the IFP precipitation (Mauk et al., 2017; Salveter et al., 2022), hence the results by Benmahi et al. (2024) should not be applied directly to the IFP. Nevertheless, this suggests that estimates of the electron energy inferred by assuming a monoenergetic energy distribution should be taken cautiously, and a model dedicated to the IFP emission in the IR band should be used, alongside the results from Juno's particle detectors. At present, such a model has not been developed yet, and therefore the comparison with the model by Tao et al. (2011) should be considered as a qualitative indication, rather than a precise constraint on the electron energy.

Regarding the timescales, it is worth noting that, for many purposes, an exponential decay is often considered a reasonable approximation. In fact, it comprehends a combination of multiple effects: electron precipitation does not cease abruptly, but rather diminishes gradually in most cases, and $H_3^+$ is distributed across different altitudes with varying recombination rates.

However, the actual temporal profile is more a hyperbolic function than an exponential one, and this becomes more evident when electron precipitation rates are variable on short timescales. Understanding this concept is crucial to avoid a common oversimplification: there is no single characteristic timescale for $H_3^+$ decay. Achilleos et al., (1998) estimated the $H_3^+$ lifetime to be from the 10s of seconds in the auroral region, to 1000 seconds in the non-auroral region; Miller et al. (2020) reports about ~100 s; various other values are reported by Stallard et al. (2002), Watanabe et al. (2018), and Mura et al. (2018): all these values do not contradict the present results, but reflect different stages (early or late) of the $H_3^+$ density decay. This is particularly important when considering comparative measurements between IR and UV emissions (Gerard et al., 2018). In fact, IR measurements present two additional complexities: i) the emission occurs at a different altitude than UV, and ii) the column density is not directly proportional to the electron flux. Instead, the relationship is non-linear, as it results from eq. 10 in Tao et al. (2011). This favors the detection of lower column densities, revealing details that may not be observable in the UV.

Finally, a remarkable aspect of this study is that we derived both the column density and the factor $f$ (which converts the column density into radiance) solely from the temporal profile of the $H_3^+$ radiance, finding very good agreement with previous literature. No additional assumptions were made, except for the value of the recombination rate $k$ from Sundström et al. (1994). This approach demonstrates the robustness of the model for $H_3^+$ decay, and highlights the unique insights gained from analyzing IR emissions in this context.

## Open Research

The JIRAM dataset used for our analysis is publicly available at the Juno Archive at the Planetary Atmospheres Node:
https://pds-atmospheres.nmsu.edu/PDS/data/PDS4/juno_jiram_bundle/data_calibrated/.
The code and the tabulated data are available at https://doi.org/10.5281/zenodo.13935220

## Acknowledgments

We thank Agenzia Spaziale Italiana (ASI) for the support of the JIRAM contribution to the Juno mission. This work is funded by the ASI–INAF Addendum n. 2016-23-H.3-2023 to grant 2016-23-H.0. This work was supported by the Fonds de la Recherche Scientifique – FNRS under Grant(s) No T003524F. V. Hue acknowledges support from the French government under the France 2030 investment plan, as part of the Initiative d'Excellence d'Aix-Marseille Université – A*MIDEX AMX-22-CPJ-04, as well as support from CNES for the Juno mission.

# Appendix

## A1 Instrument

JIRAM is both an imager and a spectrometer in the infrared spectrum (Adriani et al., 2008, 2017). Its imaging channel captures images in two spectral bands, M and L, using filters positioned over the detector. The imaging channel uses a single detector (266 x 432 pixels) divided into two segments (128 x 432 pixels each) by a non-sensitive strip, with each segment corresponding to a different filter. The L band, equipped with a band-pass filter between 3.3 and 3.6 µm, is particularly effective at capturing a significant portion of the infrared H3+ emission (lines at ~3.3, ~3.4 and ~3.55 µm), and assists in reconstructing the context of the measured spectra. JIRAM's spatial resolution is exceptionally high, with an angular resolution of approximately 0.01°. The L band's field of view (FoV) is 5.87° by 1.74°, with spatial resolution varying based on the spacecraft's distance from the surface, ranging from tens to hundreds of kilometers. The second filter, the M band (4.5 to 5 µm), is designed for mapping atmospheric thermal structures.

Images are captured with a time resolution of 1 second, with intervals of 30 seconds between consecutive images. The L band channel does experience some background noise, likely due to stray light beneath the filter or electronic bleeding within the detector. However, this noise is mitigated using the method described by Mura et al., 2017.

## A2 Data set and methods

Tables A1 gives the list of the JIRAM images used for reconstructing the vertical profile of the Io auroral footprint (Figure 1). The code for the model, and the MCMC information are available at a public repository (see Mura et al., 2024).

Tables A2 and A3 gives the list of the JIRAM images used for the time variability study. ROI for orbit 13 is located at ($X_{SIII}$=14450 km, $Y_{SIII}$=-33880 km) in the orthographic projection; ROI for orbit 26 is located at ($X_{SIII}$=15690 km, $Y_{SIII}$=20660km) in the orthographic projection. ROI for orbit 13 has been increasingly adjusted with an offset on the X axis of 7 km per observation in order to keep the sub-dot aligned with the center of the ROI. The signal in the ROI for orbit 26 has been adjusted by removing 0.04 mW m$^{-2}$ sr$^{-1}$ to account for a small background signal in the image.

Figure A1 gives the half-life as a function of the column density, as calculated by taking the model functions in Figure 3.

**Table A1. Observations used for Figure 1, orbit 31**

| Date | Time | File name in PDS |
|---|---|---|
| 2020-12-30 | 21:05:26.211 | JIR_IMG_EDR_2020365T210530_V01.IMG |
| 2020-12-30 | 21:05:56.290 | JIR_IMG_EDR_2020365T210600_V01.IMG |
| 2020-12-30 | 21:06:26.380 | JIR_IMG_EDR_2020365T210630_V01.IMG |
| 2020-12-30 | 21:06:56.470 | JIR_IMG_EDR_2020365T210700_V01.IMG |
| 2020-12-30 | 21:07:26.559 | JIR_IMG_EDR_2020365T210730_V01.IMG |
| 2020-12-30 | 21:07:56.650 | JIR_IMG_EDR_2020365T210800_V01.IMG |
| 2020-12-30 | 21:08:26.739 | JIR_IMG_EDR_2020365T210830_V01.IMG |
| 2020-12-30 | 21:08:56.831 | JIR_IMG_EDR_2020365T210900_V01.IMG |
| 2020-12-30 | 21:10:56.941 | JIR_IMG_EDR_2020365T211101_V01.IMG |
| 2020-12-30 | 21:11:27.043 | JIR_IMG_EDR_2020365T211131_V01.IMG |
| 2020-12-30 | 21:11:57.136 | JIR_IMG_EDR_2020365T211201_V01.IMG |
| 2020-12-30 | 21:12:27.229 | JIR_IMG_EDR_2020365T211231_V01.IMG |
| 2020-12-30 | 21:12:57.322 | JIR_IMG_EDR_2020365T211301_V01.IMG |
| 2020-12-30 | 21:13:27.415 | JIR_IMG_EDR_2020365T211331_V01.IMG |
| 2020-12-30 | 21:14:57.449 | JIR_IMG_EDR_2020365T211501_V01.IMG |
| 2020-12-30 | 21:15:27.552 | JIR_IMG_EDR_2020365T211531_V01.IMG |
| 2020-12-30 | 21:15:57.647 | JIR_IMG_EDR_2020365T211601_V01.IMG |
| 2020-12-30 | 21:16:27.744 | JIR_IMG_EDR_2020365T211631_V01.IMG |
| 2020-12-30 | 21:16:57.841 | JIR_IMG_EDR_2020365T211701_V01.IMG |
| 2020-12-30 | 21:17:27.939 | JIR_IMG_EDR_2020365T211732_V01.IMG |
| 2020-12-30 | 21:18:57.983 | JIR_IMG_EDR_2020365T211902_V01.IMG |
| 2020-12-30 | 21:19:28.088 | JIR_IMG_EDR_2020365T211932_V01.IMG |
| 2020-12-30 | 21:19:58.187 | JIR_IMG_EDR_2020365T212002_V01.IMG |
| 2020-12-30 | 21:20:28.289 | JIR_IMG_EDR_2020365T212032_V01.IMG |
| 2020-12-30 | 21:20:58.392 | JIR_IMG_EDR_2020365T212102_V01.IMG |
| 2020-12-30 | 21:21:28.495 | JIR_IMG_EDR_2020365T212132_V01.IMG |
| 2020-12-30 | 21:22:58.550 | JIR_IMG_EDR_2020365T212302_V01.IMG |
| 2020-12-30 | 21:23:28.658 | JIR_IMG_EDR_2020365T212332_V01.IMG |
| 2020-12-30 | 21:23:58.764 | JIR_IMG_EDR_2020365T212402_V01.IMG |
| 2020-12-30 | 21:24:28.871 | JIR_IMG_EDR_2020365T212433_V01.IMG |
| 2020-12-30 | 21:24:58.978 | JIR_IMG_EDR_2020365T212503_V01.IMG |
| 2020-12-30 | 21:25:29.088 | JIR_IMG_EDR_2020365T212533_V01.IMG |

**Table A2. Observations used for Figure 2, 3 (panel A) and 4, orbit 13.**

| Date | Time | File name in PDS | Average radiance in the ROI (mW m$^{-2}$ sr$^{-1}$) |
|---|---|---|---|
| 2018-05-24 | 07:12:27.931 | JIR_IMG_EDR_2018144T071232_V01.IMG | 0.292 |
| 2018-05-24 | 07:12:58.389 | JIR_IMG_EDR_2018144T071302_V01.IMG | 0.223 |
| 2018-05-24 | 07:13:28.848 | JIR_IMG_EDR_2018144T071333_V01.IMG | 0.157 |
| 2018-05-24 | 07:13:59.307 | JIR_IMG_EDR_2018144T071403_V01.IMG | 0.117 |
| 2018-05-24 | 07:14:29.765 | JIR_IMG_EDR_2018144T071433_V01.IMG | 0.094 |
| 2018-05-24 | 07:15:00.224 | JIR_IMG_EDR_2018144T071504_V01.IMG | 0.083 |
| 2018-05-24 | 07:15:30.682 | JIR_IMG_EDR_2018144T071534_V01.IMG | 0.082 |

**Table A3. Observations used for Figure 3 (panel B), orbit 26**

| Date | Time | File name in PDS | Average radiance in the ROI (mW m$^{-2}$ sr$^{-1}$) |
|---|---|---|---|
| 2020-04-10 | 15:51:40.795 | JIR_IMG_EDR_2020101T155144_V01.IMG | 0.207 |
| 2020-04-10 | 15:52:10.943 | JIR_IMG_EDR_2020101T155215_V01.IMG | 0.174 |
| 2020-04-10 | 15:52:41.090 | JIR_IMG_EDR_2020101T155245_V01.IMG | 0.168 |
| 2020-04-10 | 15:53:11.238 | JIR_IMG_EDR_2020101T155315_V01.IMG | 0.137 |
| 2020-04-10 | 15:53:41.384 | JIR_IMG_EDR_2020101T155345_V01.IMG | 0.117 |
| 2020-04-10 | 15:54:11.532 | JIR_IMG_EDR_2020101T155415_V01.IMG | 0.111 |
| 2020-04-10 | 15:54:41.678 | JIR_IMG_EDR_2020101T155445_V01.IMG | 0.102 |
| 2020-04-10 | 15:55:11.826 | JIR_IMG_EDR_2020101T155515_V01.IMG | 0.099 |

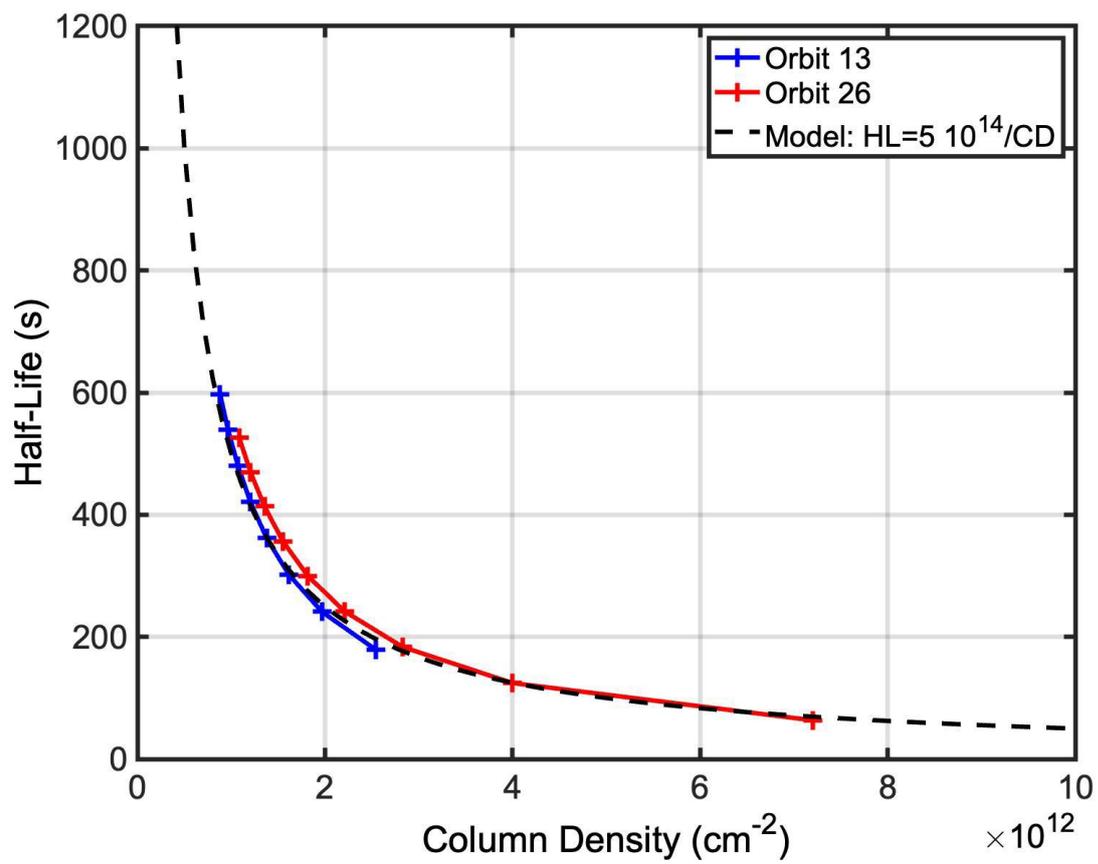

**Figure A1**. Half-life obtained from the models in Figure 3, function of the column density. Blue: orbit 13; red: orbit 26; black: analytical formula.